\documentclass[leqno, 12pt]{article}
\usepackage{times}
\usepackage[german,italian,english]{babel}
\usepackage{amssymb,amsmath}
\usepackage[latin1]{inputenc}
\usepackage[multiple,para]{footmisc}
\usepackage{cite}
\title{How Weyl stumbled across electricity while pursuing mathematical justice}
\author{Alexander Afriat}
\begin{document}
\maketitle
\begin{abstract}
\noindent It is argued that Weyl's theory of gravitation and electricity came out of `mathematical justice': out of the \emph{equal rights direction and length}. Such mathematical justice was manifestly at work in the context of discovery, and is enough (together with a couple of simple and natural operations) to derive all of source-free electromagnetism. Weyl's repeated references to coordinates and gauge are taken to express equal treatment of direction and length.
\end{abstract}
\section{Introduction}
It is almost always claimed\footnote{Certainly by Folland \cite{Folland}, Trautman \cite{Trautman}, Yang \cite{Yang}, Perlick \cite{Perlick}, Vizgin \cite{Vizgin} and others.} that Weyl \emph{deliberately} unified gravitation and electricity in the rectification of general relativity he attempted in 1918. In fact the unification, as Bergia \cite{Bergia} and Ryckman \cite{Ryckman} have pointed out and a couple of passages\footnote{\cite{Weyl1918GE}~pp.~148-9: ``\foreignlanguage{german}{Indem man die erwähnte Inkonsequenz beseitigt, kommt eine Geometrie zustande, die überraschenderweise, auf die Welt angewendet, \em nicht nur die Gravitationserscheinungen, sondern auch die des elektromagnetischen Feldes erklärt\upshape.}"}\footnote{\cite{EinsteinWeyl}: ``\foreignlanguage{german}{Übrigens müssen Sie nicht Glauben, daß ich von der Physik her dazu gekommen bin, neben der quadratische noch die lineare Differentialform in die Geometrie einzuführen; sondern ich wollte wirklich diese "`Inkonsequenz,"' die mir schon immer ein Dorn im Auge gewesen war, endlich einmal beseitigen und bemerkte dann zu meinem eigenen erstaunen: das sieht so aus, als erklärt es die Elektrizität.}"} show, was the unintended outcome of \emph{a priori}\footnote{As opposed to ``experimentally founded" or even ``empirically justified" (with respect to the past; \emph{a posteriori} justification is of course another matter). \emph{A priori} considerations can be {\ae}sthetic or mathematical, for instance.} prejudice.\footnote{I say ``prejudice"---and not ``principle" or ``assumption," for instance---to emphasize the unexpected, gratuitous, almost unaccountable character of the considerations.} But what prejudice?

The evidence as I read it suggests the theory came straight out of Weyl's sense of mathematical `justice,' which led him to put the direction and length of a vector on an equal footing. Levi-Civita \cite{Levi-Civita} had discovered that the parallel transport determined by Einstein's covariant derivative was not integrable---while length, far from depending on the path taken, remained unaltered. For Weyl this was unfair: both features deserved the same treatment.\footnote{\cite{Weyl1918GE} p.\ 148: ``\foreignlanguage{german}{[\thinspace\dots] und es ist dann von vornherein ebensowenig anzunehmen, daß das Problem der Längenübertragung von einem Punkte zu einem endlich entfernten integrabel ist, wie sich das Problem der Richtungsübertragung als integrabel herausgestellt hat.}"} He remedied with a connection that made \emph{congruent} transport (of length) just as path-dependent as parallel transport. This `total' connection restored justice through a \emph{length connection} it included, an inexact one-form Weyl couldn't help identifying with the electromagnetic four-potential $A$,\footnote{Writing the geometrical objects of which Weyl often gives no more than the components may seem anachronistic. But he undeniably sees the geometry behind the components, and sometimes explicitly refers to the underlying geometrical object, \emph{e.g}.\ ``\foreignlanguage{german}{Ein (kontravarianter) \emph{Vektor} $\mathfrak{x}$ im Punkte $P$ hat mit Bezug auf jedes Koordinatensystem gewisse $n$ Zahlen $\xi^i$, die sich beim Übergang zu einem andern Koordinatensystem [\thinspace\dots]}" (\negthinspace\cite{Weyl1918GE} p.\ 149), or ``\foreignlanguage{german}{[\thinspace\dots] die $g_{ik}$ [\thinspace\dots] bilden die Komponenten des Gravitationspotentials. [\thinspace\dots] einem Viererpotential [\thinspace\dots] dessen Komponenten $\varphi_i$ [\thinspace\dots]}" (previous page). So I see no harm in using $A$ to denote the ``Viererpotential," whose components are $A_i=\langle A,\partial_i\rangle=\varphi_i$, and $g$ to denote the ``Gravitationspotential," with components $g_{ik}=g(\partial_i,\partial_j)$; and so on.} whose four-curl $F = dA$, being closed (for $dF=d^2A$ vanishes identically), provides Maxwell's two homogeneous equations. Source-free electromagnetism (up to Hodge duality at any rate) thus came, quite unexpectedly, out of Weyl's surprising sense of mathematical justice.

Admittedly there were also intimations,\footnote{\cite{Weyl1918GE} p.\ 148, for instance: ``\foreignlanguage{german}{In der oben charakterisierten Riemannschen Geometrie hat sich nun ein letztes ferngeometrisches Element erhalten [\thinspace\dots].}" Or (same page): ``\foreignlanguage{german}{Eine Wahrhafte Nahe-Geometrie darf jedoch nur ein Prinzip der Übertragung einer Länge von einem Punkt zu einem unendlich benachbarten kennen [\thinspace\dots].}" Further adumbrations---such as the title: ``Reine Infinitesimalgeometrie"---can be found in \cite{Weyl1918RE}, which came out about half a year after the communication of \cite{Weyl1918GE}.} from the beginning, announcing a `telescepticism' (\ref{telescepticism}) opposed to distant comparisons, which Ryckman has rightly traced back to Husserl. Other texts\footnote{\cite{ZG} p.\ 125: ``\foreignlanguage{german}{Jene "`objektive"' Welt, welche die Physik aus der von uns unmittelbar erlebten Wirklichkeit herauszuschälen bestrebt ist, können wir nach ihrem bezeichenbaren Gehalt nur durch mathematische Begriffe erfassen. Um aber die Bedeutung, welche dieses mathematische Begriffssystem für die Wirklichkeit besitzt, zu kennzeichnen, müssen wir irgendwie seinen Zusammenhang mit dem unmittelbar Gegebenen zu beschreiben versuchen, eine Aufgabe der Erkenntnistheorie, die naturgemäß nicht mit physikalischen Begriffen allein, sondern nur durch beständige Berufung auf das in Bewußtsein anschaulich Erlebte geleistet werden kann}"; the introduction to \cite{RZM1}; and \cite{Kontinuum}. All three were brought to my attention by a referee.} from the same period indicate a rather murky phenomenological background that helps situate such intimations within a developing `infinitesimal' agenda. But this conceptual framework, as it appears around 1918 at any rate, is logically insufficient on its own to bring together gravitation and electricity. The later texts quoted in footnotes 27 and 28 can be taken to provide something approaching logical sufficiency, which could then be ascribed to the whole agenda \emph{a posteriori}. My interpretation (\ref{two}) of an insistent contraposition of coordinates and gauge has led me to stress the role of `mathematical justice' instead, whose compelling logical sufficiency spares one the ambiguities and chronological uncertainties of a more ambitious (and ideologically richer) kind of reconstruction. I have little or nothing to add to Ryckman's interpretation, which retains its validity and great interest. But since a viable alternative can only enrich our understanding of Weyl's theory, especially of its origins, I will contend that what was really at work in the spring of 1918, what effectively gave rise to the theory, was the \emph{equal rights of direction and length}. As the textual evidence underdetermines its interpretation, why not explore the available `freedom' and offer a new reconstruction consistent with that evidence. The freedom needn't be entirely even and uniform, without relief or texture; it could be varied, with `accidents' of all sorts; regions of it may be favoured by the familiar criteria of simplicity, elegance, economy, convenience and so forth. Needless to say I claim to have found a distinguished `sweet spot' within the freedom, and that my interpretation is not only \emph{compatible} with the evidence, but even \emph{suggested} by it---especially by Weyl's harping on coordinates \emph{vs}.\ gauge, the way I read it at any rate \ldots

\section{Background: Einstein, Levi-Civita}
We can begin with aspects of Einstein's theory of gravitation, since Weyl's theory grew out of it. What interests us above all is affine structure, given by the Christoffel symbols $\mathit{\Gamma}^a_{bc}$. Through the geodesic equation
\begin{equation}\label{one}
\frac{d^2x^a}{ds^2}+\mathit{\Gamma}^a_{bc}\frac{dx^b}{ds}\frac{dx^c}{ds}=0
\end{equation}
($a=0\textrm{,}\ldots\textrm{,}\thinspace 3$) and the wordlines satisfying it, the Christoffel symbols provide a notion of (para\-metrised\footnote{For (\ref{one}) determines an equivalence class [$s$] of affine parameters, each parameter of which gives the proper time of a regular clock, with its own zero and unit of time. The parameters belonging to [$s$] are related by affine transformations $s\mapsto \upsilon s+\zeta$, where the constants $\upsilon$ and $\zeta$ give the unit and zero. The constant $\upsilon$ is typically chosen so that $g(\partial_0,\partial_0)=1$.}) straightness, of inertial, unaccelerated motion, of free fall.

The left-hand side of (\ref{one}) gives the components $\langle dx^a,\nabla_{\dot{\sigma}}\dot{\sigma}\rangle$ of the covariant derivative $\nabla_{\dot{\sigma}}\dot{\sigma}$ of the vector $\dot{\sigma}$ with components $dx^a/ds=\langle dx^a,\dot{\sigma}\rangle$, in the direction $\dot{\sigma}$ tangent to the worldline
\begin{equation*}
\begin{split}
\sigma :\thinspace &I\rightarrow M\\
&s\mapsto\sigma\textrm{(}s\textrm{)}
\end{split}
\end{equation*}
with coordinates $\sigma^a\textrm{(}s\textrm{)}=x^a(\sigma\textrm{(}s\textrm{)})$, where $I$ is an appropriate interval and $M$ the differential manifold representing the universe; $a=0\textrm{,}\ldots\textrm{,}\thinspace 3$. The Christoffel symbols are related to $\nabla$ by
$$\mathit{\Gamma}^a_{bc}=\langle dx^a,\nabla_{\partial_b}\partial_c\rangle\textrm{,}$$
where the basis vectors $\partial_a=\partial_{a\textrm{(}x\textrm{)}}=\partial/\partial x^a$ are tangent to the coordinate lines of the system $x^a$.

Einstein only appears to have explored the \emph{infinitesimal} behaviour of the parallel transport determined by his covariant derivative. It was Levi-Civita \cite{Levi-Civita} who first understood that if $\nabla g$ vanishes, as in Einstein's theory, the direction of the vector $V_s\in T_{\sigma\textrm{(}s\textrm{)}}M$ transported according to $\nabla_{\dot{\sigma}}V_s=0$ depends\footnote{\cite{Levi-Civita} p.\ 175: ``\foreignlanguage{italian}{La direzione parallela in un punto generico $P$ ad una direzione ($\alpha$) uscente da un altro punto qualsiasi $P_0$ dipende in generale dal cammino secondo cui si passa da $P_0$ a $P$. L'indipendenza dal cammino è proprietà esclusiva delle varietà euclidee.}"} on the path $\sigma$ taken---whereas the squared length $l_s=g(V_s,V_s)$ remains constant along $\sigma$, for
$$\nabla g=0=\nabla_{\dot{\sigma}}V_s$$
means that $dl_s/ds=\nabla_{\dot{\sigma}}l_s$ vanishes.
\section{The emergence of Weyl's theory}
\subsection{The equal rights of direction and length}\label{two}
Weyl felt that as parallel transport depended on the path taken, congruent transport ought to as well. In fact his generalisation of Einstein's theory appears to have been almost entirely determined by the intention of putting direction and length on an equal footing. The following table\footnote{Parts of it were inspired by Coleman and Korté \cite{CK} pp. 204-5, 211-2.}---parts of which may for the time being be more intelligible than others---outlines Weyl's programme.
$$\begin{array}{ccc}
    \fbox{\textbf{DIRECTION}} & & \fbox{\textbf{LENGTH}}\\
    \textrm{coordinates (up to gauge)} & & \textrm{gauge}\\
    \textrm{gravitation} & & \textrm{electricity}\\
    \textrm{parallel}& \textbf{transport} &\textrm{congruent}\\
    \mathit{\Gamma}^a_{bc}\textrm{ (Levi-Civita)}& \textbf{connection}& A\\
    \delta V^a=-\mathit{\Gamma}^a_{bc}X^bV^c & &\delta l=-\langle\alpha,X\rangle=-A_b X^b l\\
    R^a_{bcd}\,(\textrm{of }\mathit{\Gamma}^a_{bc})& \textbf{curvature}& F=dA\\
    \textrm{coordinates }y^a\textrm{ (at }P\textrm{):}\thinspace\mathit{\Gamma}^a_{bc}=0 &\textbf{geodesic}&\textrm{gauge (at }P\textrm{): }A^{\prime}=A+d\lambda=0\\
    \ddot{x}^a+\mathit{\Gamma}^{a}_{bc}\dot{x}^b \dot{x}^c\mapsto \ddot{y}^a & \textbf{equiv. princ.}&\alpha=-lA \mapsto\alpha^{\prime}=0\\
 \end{array}$$
 A few words about ``coordinates (up to gauge)." The parallel between coordinates and gauge, which Weyl draws\footnote{\cite{Weyl1918GE} p.\ 150: ``\foreignlanguage{german}{Zum Zwecke der analytischen Darstellung haben wir 1.\ ein bestimmtes Koordinatensystem zu wählen und 2.\ in jedem Punkte den willkürlichen Proportionalitätsfaktor, mit welchem die $g_{ik}$ behaftet sind, festzulegen. Die auftretenden Formeln müssen dementsprechend eine doppelte Invarianz\-eigenschaft besitzen: 1.\ sie müssen \emph{invariant} sein \emph{gegenüber beliebigen stetigen Koordinatentransformationen}, 2.\ sie müssen ungeändert bleiben, \emph{wenn man die $g_{ik}$ durch $\lambda g_{ik}$ ersetzt}, wo $\lambda$ eine willkürliche stetige Ortsfunktion ist. Das Hinzutreten dieser zweiten Invarianzeigenschaft ist für unsere Theorie charakteristisch.}"}\footnote{\cite{Weyl1918RE} p.\ 396: ``\foreignlanguage{german}{Zum Zwecke der analytischen Darstellung denken wir uns 1.\ ein bestimmtes Koordinatensystem und 2.\ den an jeder Stelle willkürlich zu wählenden Proportionalitätsfaktor im skalaren Produkt festgelegt; damit ist ein "`\emph{Bezugssystem}"'$^{\textrm{9}}$ für die analytische Darstellung gewonnen.}" And footnote 9: ``\foreignlanguage{german}{Ich unterscheide also zwischen "`Koordinatensystem"' und "`Bezugssystem."'}"}\footnote{\cite{Weyl1918RE} p.\ 398: \foreignlanguage{german}{``In alle Größen oder Beziehungen, welche metrische Verhältnisse analytisch darstellen, müssen demnach die Funktionen $g_{ik}$, $\varphi_i$ in solcher Weise eingehen, daß Invarianz stattfindet 1.\ gegenüber einer beliebigen Koordinatentransformation ("`Koordinaten-Invarianz"') und 2.\ gegenüber der Ersetzung von (7) durch (8) ("`Maßstab-Invarianz"')."}}\footnote{\cite{Weyl1919} p.\ 101: ``Um den physikalischen Zustand der Welt an einer Weltstelle durch Zahlen charakterisieren zu können, muß 1.\ die Umgebung dieser Stelle auf \emph{Koordinaten} bezogen sein und müssen 2.\ gewisse \emph{Maßeinheiten} festgelegt werden. Die bisherige E\thinspace i\thinspace n\thinspace s\thinspace t\thinspace e\thinspace i\thinspace n\thinspace sche Relativitätstheorie bezieht sich nur auf den ersten Punkt, die Willkürlichkeit des Koordinatensystems; doch gilt es, eine ebenso prinzipielle Stellungnahme zu dem zweiten Punkt, der Willkürlichkeit der Maßeinheit, zu gewinnen."} over and over, can be seen as a parallel between direction and length. For surely Weyl does not mean ``coordinates \emph{including gauge}---as opposed to gauge," which would be redundant.\footnote{\emph{Cf.} \cite{Weyl1918GE} p.\ 149: ``\foreignlanguage{german}{Wird die Mannigfaltigkeit der Raumpunkte durch Koordinaten $x_i$ dargestellt, so sind durch die Metrik im Punkte $P$ die $g_{ik}$ nur ihrem Verhältnis nach festgelegt.}"} And up to gauge, coordinates provide no more than direction: The coordinates $x^a$ assign to each event $P\in M$ a basis $\partial_a\in T_PM$, and a dual basis
$$dx^a=g^{\flat}(\partial_a)=g(\partial_a,\cdot\thinspace )\in T^*_PM$$
giving the components $V^a=\langle dx^a,V\rangle$ of any vector $V\in T_PM$; $a=0\textrm{,}\ldots\textrm{,}\thinspace 3$. The recalibration\footnote{The convenient `exponential' recalibration is not used by Weyl.} $g\mapsto e^{2\lambda}g$ induces a transformation $V\mapsto e^{\lambda}V$, or $V^a\mapsto e^{\lambda}V^a$, through
$$e^{2\lambda}g(V,V)=g(e^{\lambda}V,e^{\lambda}V)=g(e^{\lambda}\partial_a,e^{\lambda}\partial_b)V^aV^b=g(\partial_a,\partial_b)e^{\lambda}V^ae^{\lambda}V^b.$$
Direction, given by the ratios
$$e^{\lambda}V^0:e^{\lambda}V^1:e^{\lambda}V^2:e^{\lambda}V^3=V^0:V^1:V^2:V^3\textrm{,}$$
remains unaffected.

Weyl clearly distinguishes between a `stretch' (like a \emph{stretch} of road) and its numerical length, determined by the gauge chosen. Just as a direction $[e^{\lambda}V]_{\textrm{(all }\lambda\textrm{)}}$ is `expanded' with respect to a coordinate system, which provides its numerical representation (the ratios $V^0:\cdots :V^3$), a stretch gets `expanded' in a gauge, which likewise gives a numerical representation, the (squared) length
$$l=e^{2\lambda}g(V,V).$$

The rest of the table should in due course become clearer. Let us now see how the inexact one-form $A$, which gives rise to so much of electromagnetism, emerges from the equal rights of direction and length.
\subsection{Electromagnetism from equal rights}\label{three}
Weyl calls a manifold $M$ \em affinely connected \upshape if the tangent space $T_PM$ at every point $P\in M$ is connected to all the neighbouring tangent spaces $T_{P^{\prime}}M$ by a mapping
\begin{equation*}
\begin{split}
\mathit{\Xi}_X:\thinspace\thinspace &T_PM\rightarrow T_{P^{\prime}}M\\
&V_P\mapsto V_{P^{\prime}}=\mathit{\Xi}_XV_P
\end{split}
\end{equation*}
linear both in the `main' argument $V_P\in T_PM$ and in the (short\footnote{The necessary shortness of $X$ seems inconsistent with linearity, which would `connect' $P$ with the entire tangent space $T_PM$ and not just with the small neighbourhood `covering' $M$. In this context it may be best to view the linearity in the directional argument as being appropriately restricted (of course the length of $X$ does not matter in differentiation, in which limits are taken).}) directional argument $X=P^{\prime}-P$, where $P^{\prime}$ (being near $P$) and hence $X$ are viewed as lying in $T_PM$. Being linear, $\mathit{\Xi}_X$ will be represented by a matrix:
\begin{equation*}
\begin{split}
\mathit{\Xi}^a_c=\langle dx^a,\mathit{\Xi}_X\partial_c\rangle=\mathit{\Xi}^a_{bc}X^b\\
=\langle dx^a,\mathit{\Xi}_{\partial_b}\partial_c\rangle\langle dx^b,X\rangle .
\end{split}
\end{equation*}
Weyl specifically refers to the components $\delta V^a=\langle dx^a_{P^{\prime}}, V_{P^{\prime}}\rangle -\langle dx^a_P,V_P\rangle$, requiring them to be linear in the components $X^b$ and $V^c_P=\langle dx^c_P,V_P\rangle$. The bilinear function
$$\mathit{\Gamma}^a(\{X^b\},\{V^c\})=\delta V^a$$
will be a matrix, represented by $\mathit{\Gamma}^a_{bc}$; the difference $\delta V^a$ is therefore $-\mathit{\Gamma}^a_{bc}X^bV^c.$

With respect to the \emph{geodesic} coordinates $y^a$ which make
\begin{equation*}\begin{split}
\mathit{\Gamma}^a_c&=\mathit{\Gamma}^a_{bc}X^b\\
&=\langle dy^a,\nabla_X\partial_{c\textrm{(}y\textrm{)}}\rangle
\end{split}\end{equation*}
and $\delta V^a$ vanish, leaving the components $V^a$ unchanged, $\mathit{\Xi}^a_c$ becomes the identity matrix
$$\delta^a_c=\langle dy^a,\mathit{\Xi}_X\partial_{c\textrm{(}y\textrm{)}}\rangle\leftrightarrow\textrm{diag}\textrm{(}1,1,1,1\textrm{)}.$$
Physically this has to do with the equivalence principle, according to which a gravitational field $\mathit{\Gamma}^a_{bc}$ can always be eliminated or generated at $P$ by an appropriate choice of coordinates.

With equal rights in mind Weyl turns to length, using the very same scheme. To clarify his procedure we can take just a single component of the difference $\{\delta V^0,\ldots,\delta V^3\}$, calling it $\delta l$ (this will be the `squared-length-difference scalar').\footnote{Weyl appears to use $d$ and $\delta$ interchangeably, and $d$ in a way---see footnote 23---that is unusual not only today, but even then. He does not distinguish between the \emph{scalar} representing the difference in squared length, and the corresponding \emph{one-form} (as we would call it); but the distinction nonetheless seems useful.} The upper index of $\mathit{\Gamma}^a_{bc}$ accordingly disappears, leaving\footnote{We can perhaps think of the hybrid, intermediate connection $\mathit{\Gamma}_{bc}$ as being something like $\langle A,\nabla_{\partial_b}\partial_c\rangle$.}
$$\delta l=\mathit{\Gamma}_{bc}X^bV^c.$$
If we now take a single component of the main argument $\{V^0,\ldots ,V^3\}$, calling it $l$ (this will be the squared length), the second index of $\mathit{\Gamma}_{bc}$ disappears as well, and we are left with
$$\delta l=\mathit{\Gamma}_bX^bl\textrm{,}$$
where $\mathit{\Gamma}_b=\langle A,\partial_b\rangle$ are the components of a one-form,\footnote{One may wonder how the tensor $A$ can be the counterpart of the connection $\mathit{\Gamma}^a_{bc}$, which is not a tensor. The components $A_a=\langle A,\partial_a\rangle =\mathit{\Gamma}_a$ only transform as a tensor with respect to coordinate transformations $A_a\mapsto\bar{A}_b=A_a\langle d\bar{x}^b,\partial_{a\textrm{(}x\textrm{)}}\rangle$, however; with respect to recalibration $A_a\mapsto A^{\prime}_a=A_a+\partial_a\lambda$ the components $A_a$ do not transform `tensorially,' and can be locally cancelled, for instance.} denoted $A$ with electricity in mind.

 But this is not really Weyl's argument, which is better rendered as follows. The object $A$ generating the squared-length-difference scalar $\delta l$ has to be linear in the squared length $l$ and the direction $X$. A linear function $A\textrm{(}l,X\textrm{)}=\delta l$ of a scalar $l$ and vector $X$ yielding a scalar $\delta l$ will be a one-form:\footnote{Weyl in fact writes $dl=-ld\varphi$, whereas I write $\alpha=-lA.$ The misleading $d$'s cannot be understood globally---or even locally, in the theory of gravitation and electricity, in which $F=d^2\varphi$ will be the Faraday two-form: where $d\varphi$ is closed, in other words the differential (even only locally) of a function $\varphi$, there would be no electromagnetism.}
\begin{equation*}
 \begin{split}
 \delta l
& =-\langle\alpha ,X\rangle =-\langle\alpha ,\partial_b\rangle\langle dx^b,X\rangle=-\alpha_bX^b\\
&=-\langle A,X\rangle l =-\langle A,\partial_b\rangle\langle dx^b,X\rangle l =-A_bX^bl\textrm{,}
 \end{split}
\end{equation*}
 where $\alpha$ is the squared-length-difference one-form. An exact one-form $A=d\mu$ would make congruent transfer integrable, removing the dependence of the recalibration
 $$e^{\int_{\gamma}A}=e^{\int d\mu}=e^{\Delta\mu}$$
on the path $\gamma: \textrm{[}0,1\textrm{]}\rightarrow M$, where $\Delta \mu=\mu_1 - \mu_0$ is the difference between the values $\mu_1=\mu (P_1)$ and $\mu_0=\mu (P_0)$ of $\mu$ at $P_1=\gamma \textrm{(}1\textrm{)}$ and $P_0=\gamma \textrm{(}0\textrm{)}$. Mathematical justice therefore demands that $A$ be inexact; so the curl $F=dA$ cannot vanish identically.

Confirmation that $A$ has to be one-form, possibly inexact, is provided by Weyl's requirement that the squared-length-difference one-form $\alpha =-lA$ be eliminable at any point $P$ by recalibration.\footnote{\cite{RZM} p.\ 122: ``\foreignlanguage{german}{Ein Punkt $P$ hängt also mit seiner Umgebung metrisch zusammen, wenn von jeder Strecke in $P$ feststeht, welche Strecke aus ihr durch kongruente Verpflanzung von $P$ nach dem beliebigen zu $P$ unendlich benachbarten Punkte $P^{\prime}$ hervorgeht. Die einzige Forderung, welche wir an diesen Begriff stellen (zugleich die weitgehendste, die überhaupt möglich ist), ist diese: Die Umgebung von $P$ läßt sich so eichen, daß die Maßzahl einer jeden Strecke in $P$ durch kongruente Verpflanzung nach den unendlich benachbarten Punkten keine Änderung erleidet.}"} As $l$ is given (and does not vanish), this amounts to
$$A+d\lambda=0$$
at $P$, where the gauge $\lambda$ is \emph{geodesic}.\footnote{By analogy one might even call it `inertial' or `unaccelerated.'} Since $d\lambda$ is a one-form, $A$ must be one too. Though $d\lambda$ is exact, Weyl only asks that it cancel $A$ \emph{at} $P$---so $A$ needn't even be \emph{closed}, or locally exact.

With $F=dA$ and its consequence $dF=0$ before him Weyl could not help seeing the electromagnetic four-potential $A$, the Faraday two-form $F=dA$ (which vanishes wherever $A$ is closed) and Maxwell's two homogeneous equations,\footnote{In full, $\nabla\cdot B=0$ and $\nabla\times E+\partial B/\partial t=0$.} expressed by $dF=0$---not to mention an electromagnetic `equivalence principle'\footnote{Lyre \cite{Lyre} speaks of a \emph{generalised equivalence principle}.} according to which the squared-length-difference scalar $\delta l$ and one-form $\alpha$, as well as the electromagnetic four-potential $A$, can be eliminated or produced at a point by an appropriate gauge function $\lambda$.

In coordinates
\begin{equation*}
F_{ab}=\thinspace F(\partial_a,\partial_b)=\partial_a A_b-\partial_b A_a\leftrightarrow \left(
                                                                               \begin{array}{cccc}
                                                                                 0 & -E_x & -E_y & -E_z \\
                                                                                 E_x & 0 & B_z & -B_y \\
                                                                                 E_y & -B_z & 0 & B_x \\
                                                                                 E_z & B_y & -B_x & 0 \\
                                                                               \end{array}
                                                                             \right)\textrm{,}
\end{equation*}
where $E_x$, $E_y$, $E_z$ are the components of the electric field and $B_x$, $B_y$, $B_z$ those of the magnetic field. Or
$F=F_{ab}\thinspace dx^a \wedge dx^b/2.$ The vanishing three-form
\begin{equation*}
\begin{split}
dF\thinspace &=\frac{1}{2}dF_{bc}\wedge dx^b \wedge dx^c\\
&=\frac{1}{6}\partial_a F_{bc}\thinspace dx^a\wedge dx^b \wedge dx^c
\end{split}
\end{equation*}
has components $dF(\partial_a,\partial_b,\partial_c)=\partial_aF_{bc}+\partial_bF_{ca}+\partial_cF_{ab}$.

Maxwell's other two equations are obtained, in `source-free' form, by setting $d^*\negthinspace F$ equal to zero, where $^*\negthinspace F$ is the Hodge dual of the Faraday two-form, with coordinates
$$(^*\negthinspace F)_{ab}=(^*\negthinspace F)(\partial_a,\partial_b)\leftrightarrow \left(
                                                                               \begin{array}{cccc}
                                                                                 0 & B_x & B_y & B_z \\
                                                                                 -B_x & 0 & E_z & -E_y \\
                                                                                 -B_y & -E_z & 0 & E_x \\
                                                                                 -B_z & E_y & -E_x & 0 \\
                                                                               \end{array}
                                                                             \right)\textrm{.}$$

Electromagnetism thus emerged, altogether unexpectedly, from the equal rights of direction and length.
\subsection{The illegitimacy of distant comparisons}\label{telescepticism}
Weyl has another \em a priori \upshape prejudice, rooted, as Ryckman \cite{Ryckman} has cogently argued, in Husserl's transcendental phenomenology. It is expressed in two similar passages,\footnote{\cite{PdMuN} p.\ 98: ``\foreignlanguage{german}{Erkennt man neben dem physischen einen Anschauungsraum an und behauptet von ihm, daß seine Maßstruktur aus Wesensgründen die euklidischen Gesetze erfülle, so steht dies mit der Physik nicht in Widerspruch, sofern sie an der euklidischen Beschaffenheit der \,u\,n\,e\,n\,d\,l\,i\,c\,h\,k\,l\,e\,i\,n\,e\,n\, Umgebung eines Punktes $O$ (in dem sich das Ich momentan befindet) festhält [\thinspace\dots]. Aber man muß dann zugeben, daß die Beziehung des Anschauungsraumes auf den physischen um so vager wird, je weiter man sich vom Ichzentrum entfernt. Er ist einer Tangentenebene zu vergleichen, die im Punkte $O$ an eine krumme Fläche, den physischen Raum, gelegt ist: in der unmittelbaren Umgebung von $O$ decken sich beide, aber je weiter man sich von $O$ entfernt, um so willkürlicher wird die Fortsetzung dieser Deckbeziehung zu einer eineindeutigen [\emph{sic}] Korrespondenz zwischen Ebene und Fläche.}"}\footnote{\cite{Weyl1931} p.\ 52: ``\foreignlanguage{german}{Die Philosophen mögen recht haben, daß unser Anschauungsraum, gleichgültig, was die physikalische Erfahrung sagt, euklidische Struktur trägt. Nur bestehe ich allerdings dann darauf, daß zu diesem Anschauungsraum das Ich-Zentrum gehört und daß die Koinzidenz, die Beziehung des Anschauungsraumes auf den physischen um so vager wird, je weiter man sich vom Ich-Zentrum entfernt. In der theoretischen Konstruktion spiegelt sich das wider in dem Verhältnis zwischen der krümmen Fläche und ihrer Tangentenebene im Punkte $P$: beide decken sich in der unmittelbaren Umgebung des Zentrums $P$, aber je weiter man sich von $P$ entfernt, um so willkürlicher wird die Fortsetzung dieser Deckbeziehung zu einer eindeutigen Korrespondenz zwischen Fläche und Ebene.}"} which roughly say: As the curvature $R$($P$) is subtle and hard to perceive directly, a ``cognizing ego" at the ``ego center" $P\in M$ takes itself to be immersed in the `psychologically privileged' tangent space $T_PM$. The universe $M$ resembles $T_PM$ in the immediate vicinity $\mathfrak{U}$ of $P$, where they practically coincide, and `cover' one another. Beyond $\mathfrak{U}$ the relation between $M$ and the `intuitive' space $T_PM$ grows looser, as the universe goes its own way, bending as the energy-momentum tensor $T$ varies.

Ryckman writes (p.\ 148) that
\begin{quote}
Weyl restricted the homogeneous space of phenomenological intuition, the locus of phenomenological \emph{Evidenz}, to what is given at, or neighboring, the cognizing ego [\thinspace\dots]. But in any case, by delimiting what Husserl termed ``the sharply illuminated circle of perfect givenness," the domain of ``eidetic vision," to the infinitely small homogeneous space of intuition surrounding the ``ego-center" [\thinspace\dots]
\end{quote}
This restriction or delimitation can be understood in two ways: directly, in terms of the limitations of our senses, and of an accordingly circumscribed domain of sensory access, of ``eidetic vision"; or more mathematically, as follows: The cognizing ego attaches a kind of intuitive `certainty' to all of $T_PM$, which, being flat and homogeneous,\footnote{Curvature (which vanishes identically) and the metric are constant.} can be captured or `understood' in its entirety once any little piece is. The universe shares that certainty as long as it resembles $T_PM$, and hence only in $\mathfrak{U}$, outside of which it is subject to all sorts of unforeseeable variations.

Integrable congruent propagation had to be rejected as allowing the certain comparison of lengths well beyond $\mathfrak{U}$, indeed at any distance, without the welcome ambiguities related to the path followed. Returning to Ryckman (p.\ 149):
\begin{quote}
Guided by the phenomenological methods of ``eidetic insight" and ``eidetic analysis", the epistemologically privileged purely infinitesimal comparison relations of \emph{parallel transport} of a vector, and the \emph{congruent displacement} of vector magnitude, will be the foundation stones of Weyl's reconstruction. The task of comprehending ``the sense and justification" of the mathematical structures of classical field theory is accordingly to be addressed through a construction or \emph{constitution} of the latter within a world geometry entirely built up from these basic geometrical relations immediately evident within a purely infinitesimal space of intuition. A wholly \emph{epistemological} project, it nonetheless coincides with the explicitly \emph{metaphysical} aspirations of Leibniz and Riemann to ``understand the world from its behaviour in the infinitesimally small."
\end{quote}
\subsection{The two prejudices}
Removed from the context of Weyl's theory, the two prejudices are entirely distinct. While one is markedly infinitesimal, the other---`mathematical justice'---has nothing (necessarily) infinitesimal about it: in a spirit of equal rights one could require, for instance, both the directions and lengths of the vectors in some set to have the same kind of distribution---uniform, say, or Gaussian---around a given vector. Nothing infinitesimal about that.

An abundant insistence in the early going on the equal rights of direction and length, together with the absence, back then, of any explicit, articulated expression of the telescepticism of \ref{telescepticism}, suggests the following account. First there was mathematical justice, which, far from being at odds with Weyl's nascent infinitesimal agenda, supported it, perhaps even suggesting aspects. In due course Weyl's `purely infinitesimal geometry' acquired more explicit transcendental-phenomen\-ological grounding (footnotes 27 and 28), which can in hindsight make the apparently gratuitous early insistence on equal rights somewhat less surprising.
\section{Compensating transformations}\label{comp}
We have seen how Weyl's theory, building on general relativity, came out of the inexact one-form $A$---whose transformations
\begin{equation}\label{due}
A \mapsto A^{\prime} = A+d\mu
\end{equation}
are counterbalanced in the theory by
\begin{equation}\label{conforme}
g\mapsto g^{\prime} = e^{\mu}g\textrm{,}
\end{equation}
leaving length unaltered. Such compensation is fundamental enough to be worth looking at briefly.

Freedom to transform $A$ according to (\ref{due}) is left by the length curvature $F=dA$, which is indifferent to an exact term $d\mu$, as
$$F=dA^{\prime}=dA+d^2\mu=dA.$$
But (\ref{due}) does change length. Transporting the vector $X_0$ from point $P_0$ with value $\mu_0 =\mu(P_0)$ to point $P_1$ with value $\mu_1=\mu(P_1)$, the final squared length $g_1(X_1,X_1)$ acquires the additional (integrable) factor $e^{\Delta \mu}$, where $\Delta \mu=\mu_1 - \mu_0$. For $\mu$ recalibrates, along a curve $\gamma$, according to
$$e^{\int_\gamma A} \mapsto e^{\int_\gamma {A^{\prime}}} = e^{\int_\gamma A} e^{\Delta \mu}\neq e^{\int_\gamma A}\textrm{,}$$
and therefore $$g_1(X_1,X_1)=e^{\int_{\gamma}A}g_0(X_0,X_0)\neq
e^{\int_{\gamma}A^{\prime}}g_0(X_0,X_0).$$
But the conformal transformation (\ref{conforme}) compensates, leaving length unchanged:
$$g_1^{\prime}(X_1,X_1)=e^{\mu_1}g_1(X_1,X_1)=e^{\int_{\gamma}A^{\prime}}g_0^{\prime}(X_0,X_0)=e^{\int_{\gamma}
A}e^{\Delta\mu}e^{\mu_0}g_0(X_0,X_0)$$
The exponents cancel, yielding the original dilation
$$ g_1(X_1,X_1)=e^{\int_{\gamma}A}g_0(X_0,X_0).$$

The metric $g$ is \emph{compatible} with the covariant derivative $\nabla$ if $\nabla g$ vanishes, in which case the straightest worldlines (satisfying $\nabla_{\dot{\sigma}}\dot{\sigma}=0$) will also be stationary, satisfying
$$\delta\negthinspace\negthinspace \int\negthinspace\negthinspace\sqrt{g\textrm{(}\dot{\sigma},\dot{\sigma}\textrm{)}}\thinspace ds
=\delta \negthinspace\negthinspace \int\negthinspace ds=0$$
too. The covariant derivative of the recalibrated metric $g^{\prime}$ only vanishes if $\mu$ is a constant (for then $d\mu$ vanishes); otherwise
$$\nabla g^{\prime}=d\mu \otimes g^{\prime}\textrm{,}$$
which combines (\ref{due}) and (\ref{conforme}), to express the weaker \emph{Weyl compatibility}.
\section{Einstein's objection}
Out of a sense of mathematical justice, then, Weyl made congruent displacement just as path-dependent as parallel transport. But experience, objected Einstein, is unfair, showing congruent dispacement to be integrable. In a letter to Weyl dated April 15th (1918) he argued\footnote{``\foreignlanguage{german}{So schön Ihre Gedanke ist, muss ich doch offen sagen, dass es nach meiner Ansicht ausgeschlossen ist, dass die Theorie die Natur entspricht. Das $ds$ selbst hat nämlich reale Bedeutung. Denken Sie sich zwei Uhren, die relativ zueinander ruhend neben einander gleich rasch gehen. Werden sie voneinander getrennt, in beliebiger Weise bewegt und dann wieder zusammen gebracht, so werden sie wieder gleich (rasch) gehen, d. h. ihr relativer Gang hängt nicht von der Vorgeschichte ab. Denke ich mir zwei Punkte $P_1$ \& $P_2$ die durch eine Zeitartige Linie verbunden werden können. Die an $P_1$ \& $P_2$ anliegenden zeitartigen Elemente $ds_1$ und $ds_2$ können dann durch mehrere zeitartigen Linien verbunden werden, auf denen sie liegen. Auf diesen laufende Uhren werden ein Verhältnis $ds_1:ds_2$ liefern, welches von der Wahl der verbindenden Kurven unabhängig ist.---Lässt man den Zusammenhang des $ds$ mit Massstab- und Uhr-Messungen fallen, so verliert die Rel.\ Theorie überhaupt ihre empirische Basis.}"} that clocks running at the same rate at one point will \emph{continue} to run at the same rate at another, however they get there---whatever the requirements of mathematical justice. Four days later he reformulated\footnote{``\foreignlanguage{german}{[\thinspace\dots] wenn die Länge eines Einheitsmassstabes (bezw. die Gang-Geschwindigkeit einer Einheitsuhr) von der Vorgeschichte abhingen. Wäre dies in der Natur wirklich so, dann könnte es nicht chemische Elemente mit Spektrallinien von bestimmter Frequenz geben, sondern es müsste die relative Frequenz zweier (räumlich benachbarter) Atome der gleichen Art im Allgemeinen verschieden sein. Da dies nicht der Fall ist, scheint mir die Grundhypothese der Theorie leider nicht annehmbar, deren Tiefe und Kühnheit aber jeden Leser mit Bewunderung erfüllen muss.}"} the objection in terms of the `proper frequencies' of atoms (rather than genuine macroscopic clocks) ``of the same sort": if such frequencies depended on the path followed, and hence on the different (electromagnetic) vicissitudes of the atoms, the chemical elements they would make up if brought together would not have the clean spectral lines one sees.

But even if experience shows congruent displacement to be integrable, it would be wrong to conclude that the equal rights of direction and length led nowhere; for the structure that came out of Weyl's surprising sense of mathematical justice would survive in our standard gauge theories, whose accuracy is less doubtful.

\section{Final remarks}
There are various levels of `experience,' ranging from the most concrete to the most abstract: from the most obvious experimental level, having to do with the results of particular experiments, to principles, perhaps even instincts, distilled from a lifetime of experience. One such principle could be Einstein's ``I am convinced that God does not play dice," to which, having---we may conjecture---noticed that the causal regularities behind apparent randomness eventually tend to emerge, he may ultimately have been led by experience: by his own direct experience, together with his general knowledge of science and the world. One would nonetheless hesitate to view so general and abstract a principle as being \emph{a posteriori}, empirical. It is clearly not \emph{a posteriori} with respect to any particular experiment; only, if at all, with respect to a very loose, general and subjective kind of ongoing experience, capable of being interpreted in very different ways.

An unexpected empirical fertility of apparently \emph{a priori} and unempirical prejudice can sometimes be accounted for in terms of a derivation, however indirect, from experience: by attributing remote empirical roots to considerations which at first seem to have nothing at all to do with experience. The world can admittedly be experienced in very different ways, some much less obvious and straightforward than others; but here we have a prejudice which---however subtle and developed one's faculties for interpreting experience---seems to be completely unempirical. Perhaps the empirical shortcomings of the theory are best blamed, then, on the totally unempirical nature of the prejudice from which it stemmed.

Or is it so completely unempirical? As mathematical justice is at issue, the principle of sufficient reason can come to mind: if there is an imbalance, an unexpected difference, there had better be a reason for it---failing which, balance, or rather justice should prevail. Even Einstein's dice may come to mind: If a situation of apparent balance, such as
\begin{equation}\label{singlet}
|\psi\rangle =\frac{1}{\sqrt 2}(|\alpha\rangle + |\beta\rangle)\textrm{,}
\end{equation}
gives rise to an imbalance (as it must, if a measurement is made), such as the eigenvalue $+1$ of the operator $A=|\alpha\rangle\langle\alpha |-|\beta\rangle\langle\beta |$, there ought to be a reason: a circumstance unrepresented in (\ref{singlet}) which favours $|\alpha\rangle $. For God does not play dice: symmetry-breaking is never \emph{entirely} spontaneous. But the `balance' before the disruption is not always so easily seen; what tells us in general which objects or entities are to be put on an equal footing, for imbalances to be visible? \emph{Judgment}, surely; a judgment somehow \emph{founded in experience}, which assesses the relevant peculiarities of the context and determines accordingly. And here Weyl's judgment and sense of balance accord the same status to direction and length---for he sees nothing to justify a preference, an injustice.

Can the success of modern gauge theories really be attributed to Weyl's sense of mathematical justice? Or is the connection between those theories and the equal rights of direction and length too tenuous to be worth speaking of? The lineage is unmistakable, and can be traced through Yang and Mills~\cite{YangMills} and Weyl~\cite{Weyl1929EG}, back to 1918; the scheme of compensation outlined in Section \ref{comp} survives in today's theories, and is central to their success \ldots\thinspace but any attempt to answer these questions would take us too far from our subject.

Whatever the relationship between mathematical justice and experience, we have a surprising example of how directly an elaborate theory can emerge from simple \emph{a priori} prejudice. The prejudice seems gratuitous in the context of discovery, and only acquires justification and phenomenological grounding years later, in an explicit, articulated `telescepticism' which provides epistemology and motivation.
\\

\noindent
I thank Ermenegildo Caccese, Dennis Dieks, John Earman, Rossella Lupacchini, Antonio Masiello, George Sparling for many fruitful discussions; the Center for philosophy of science, University of Pittsburgh, where as Visiting Fellow I began work on Weyl's theory; and above all Thomas Ryckman, for inspiration, ideas and useful criticism.

\end{document}